\documentclass[conference,usenatbib]{basi}
\usepackage[T1]{fontenc}
\usepackage[british]{babel}
\usepackage[varg]{txfonts}
\usepackage{rotating}
\usepackage{dcolumn}
\begin{document}
\title[Galactic Black Hole Sources]{Observing Galactic Black Hole Sources in Hard X-rays}
\author[A.~R.~Rao]%
       {A.~R.~Rao$^1$\thanks{email: \texttt{arrao@tifr.res.in}}\\
       $^1$Tata Institute of Fundamental Research, Mumbai, India \\
       }

\pubyear{2013}
\volume{**}
\pagerange{**--**}
%\pagerange{\pageref{firstpage}--\pageref{lastpage}}
%\status{submitted}

\date{Received --- ; accepted ---}

\maketitle
%------------------------------------------------------------------------------%
% abstract and keywords                                                        %
%------------------------------------------------------------------------------%
\label{firstpage}

\begin{abstract}
Observations of Galactic black hole sources are traditionally done in the classical X-ray range
(2 -- 10 keV) due to sensitivity constraints. Most of the accretion power, however, is
radiated above 10 keV and the study of these sources in hard X-rays has the
potential to unravel the radiation mechanisms operating at the inner region of the
accretion disk, which is believed to be the seat of a myriad of fascinating features 
like jet emission, high frequency QPO emission etc. I will briefly summarise the long
term hard X-ray observational features like spectral state identification, state
 transitions and hints of polarised emission, and describe the new insights that would be 
provided by the forthcoming
Astrosat satellite, particularly emphasising the contributions expected from the 
CZT-Imager payload.  
\\[6pt]
%
%\hbox to 30pt{\hfil}\verb|http://www.ncra.tifr.res.in/~basi/|
%
\end{abstract}

\begin{keywords}
    black holes: accretion -- observations: technique -- X-rays: stars 
\end{keywords}

%------------------------------------------------------------------------------%
% main text of the paper, using \section, \subsection, \subsubsection          %
%------------------------------------------------------------------------------%
\section{Introduction}\label{s:intro}

Though Cygnus X-1 was suggested to be a black hole `candidate' source in the early seventies 
(Webseter \& Murdin 1972),
it carried the suffix `candidate' upto the late nineties. It is only during the 
past two decades, particularly during the  `RXTE era', that we have firm identifications
of several black hole sources. The All Sky Monitor (ASM) onboard  RXTE (Levine et al. 1996) quickly identified 
several soft X-ray transients (and made the data publicly available) and the quick 
manoeuvring  capabilities of the pointed instruments PCA and HEXTE (Jahoda et al. 2006; Rothschild et al. 2008) ensured that
extensive X-ray  spectro-temporal observations are available for a large number
of objects. Further, quick optical follow up observations using the ground based optical
telescopes provided radial velocity measurements for these transients (Ozel  et al 2010 and references therein). The fact
that many black hole sources are low mass X-ray binaries with less than 1 solar mass 
for the optical companion, did not leave much uncertainties in the estimation of the mass
function and the derivation of the masses of the compact objects. Now we have secure measurements of
the masses of about 17 black hole candidate sources. A compilation of these sources is given
in Table~1.

  One of the lasting legacies of RXTE is the wide X-ray coverage for a large number of
sources (see Remillard \& McClintock 2006 for a comprehensive review). The early observations
suggested systematic variations of emission properties enabling a data driven definition
of spectral states which took into account the spectral shape as well as the variability characteristics.
The mining of the vast RXTE archives is still an ongoing process and what it has 
firmly established in the X-ray domain is the systematic variations in the spectral states
in the X-ray transients. The accompanying multi-wavelength observations also identified 
the occurrence of jet emission, coupled to the X-ray states. This has lead to a comprehensive
picture of the black hole accretion in terms of emission from the various putative emission regions.

%  \newcolumntype{d}[1]{D{.}{.}{#1}}
%\newcolumntype{}{}{}{}{}
  \begin{table}
    \caption{Black Hole Masses}\label{tab:mass}
    \medskip
    \begin{center}
      \begin{tabular}{llllc}\hline
        Source name  & \multicolumn{1}{c}{P$_{orbit}$ (days)}
                   & \multicolumn{1}{c}{M$_{BH}$ (M$_\odot$)}
 & \multicolumn{1}{c}{D(kpc)}
 & \multicolumn{1}{c}{References$^\mathrm{a}$}\\\hline
  A0620-003 &   0.33     &  6.6$\pm$0.25 & 1.06$\pm$0.12 & 1  \\
  4U~1543-47 &  1.12     &  9.4$\pm$1.0  & 7.5$\pm$0.5   & 1  \\
XTE~J1550-564&  1.54     &  9.1$\pm$0.6  & 4.4$\pm$0.5    & 1  \\
GRO~J1655-40 &  2.62     &  6.3$\pm$0.27 & 3.2$\pm$0.5   & 1  \\
V4641~Sgr    &  2.82     &  7.1$\pm$0.3  & 9.9$\pm$2.4   & 1 \\
GS~2023+338  &  6.47     & 12$\pm$2      & 2.39$\pm$0.14 & 1 \\
M33~X7       &  3.45     & 15.65$\pm$1.45&               & 1 \\
LMC~X-1      &  3.91     & 10.91$\pm$1.54&               & 1 \\
Cyg~X-1      &  5.60     & 14.8$\pm$1.0  &               & 2 \\
GRS~1915+105 & 33.5      &  14$\pm$4     &    9$\pm$3    & 3 \\
LMC~X-3      & 1.704     &  7.6$\pm$1.3  &               & 3 \\
H1705-250    & 0.520     &  6$\pm$2      &    8.6$\pm$2.1& 3 \\
GS~1124-684  & 0.433     &  7.0$\pm$0.6  & 5.89$\pm$0.26 & 3 \\
GS~2000+250  & 0.345     &  7.5$\pm$0.3  & 2.7$\pm$0.7   & 3 \\
GRS~1009-45  & 0.283     &  5.2$\pm$0.6  & 3.82$\pm$0.27 & 3 \\
GRO~J0422+32 & 0.212     &  4$\pm$1      &  2$\pm$1      & 3 \\
XTE~J1118+480& 0.171     &  6.8$\pm$0.4  &  1.7$\pm$0.1  & 3 \\\hline
               \end{tabular}\\[5pt]
      \begin{minipage}{7cm}
        \small Notes: (a) References: 1. Ozel et al. (2010); 2. Orosz et al. (2011)
3. Casares 2006.
      \end{minipage}
    \end{center}
  \end{table}

\section{The Current Accretion  Paradigm}

 The current observational paradigm for black hole accretion runs as follows
(see Zhang 2013 and  Fender \& Belloni 2012 for comprehensive reviews). At very low accretion rates,
the sources are in a hard state with a hard  spectrum and high variability. Flat spectrum radio emission,
coming from a core jet, is seen in this state  and the X-ray and radio emissions are strongly
correlated (whether the extremely faint quiescent stages have a  similar spectral state is probably still an open
question - see Plotkin et al. 2013). During the onset of an X-ray transient, presumably due to an increase in the accretion rate
due to accretion instability, the source intensity increases with the spectral shape remaining 
roughly the same. This shows a vertical path in the hardness intensity diagram and a diagonal 
path in the variability-intensity diagram. The spectral and timing properties drastically change
and the source enters into a Hard Intermediate State (HIMS). A myriad of observational episodes 
occur in this state like transition into different QPOs, superluminal jet emission etc. The source
can exhibit multiple crossing and can settle into a soft state and return to quiescence via a 
Soft Intermediate State (SIMS). One other characteristics is the cessation of jest and the 
onset of strong winds in the soft state. The transition luminosity and the peak soft X-ray luminosity
and the rate of change in luminosity are strongly correlated (Yu et al. 2004; Yu \& Zhen 2009).
 The short lived transitions show different jet emission mechanisms and these can be broadly classified
into core jet, episodic jet emission and sumperluminal jet emission. 
  The variability and QPO characteristics show a definite pattern and by associating the characteristics
QPO frequency to some length scale in the accretion disc, it is easy to visualise  a specific 
accretion disc radius. This can be characterised as a truncation radius responsible for various
activities like QPO generation, jet launching etc. Such well established patterns in the emission
characteristics prompted observers to define a 
``small number of states and their
 association with jets providing a 
good frame work to base theoretical
 studies" (Belloni et al, 2011).

\subsection{Theoretical considerations}

Most of the observational features can be coupled to the current ideas of
accretion physics. The Shakura-Sunyaev (SS) disk  (Shakura \& Sunyaev 1973) is extensively used to understand the
thermal emission. The non-thermal emission is identified with the various 
forms of inverse Compton scattering (Done et al.  2007).
In this sort of theoretical paradigms, each observations is identified with a perceived
emission region/ mechanism.

Chakrabarti and his collaborators, on the other hand, assumed that the accretion can have
two components, a SS disk coexisting with a sub-Keplerian disk (Chakrabarti \& 
Titarchuk 1995). This formalism
enabled them to avoid prescribing viscosity, but the accretion rate, however, was parameterised
with two components: a disk accretion rate and a halo accretion rate. By this method, the
transition to the inner regions could be connected through a Centrifugal pressure dominated
boundary layer (CENBOL) and the flow beyond the CENBOL could be studied through imvicid
transonic flow formalism (Chakrabarti  1989). Existence of shocks could be proved for certain flow parameters.
Though this model has several attractive features like explaining the spectral states and their
association with the QPOs (Chakrabarti et al. 2008; Nandi et al. 2012; Debnath et al. 2013), it is still to be widely accepted in the community.

% and it still
%does not provide any strong predictable observational features.

% and hence, it lacks real predictive power. The fundamental
%problems with the current ideas of accretion theories can be summarised, in an
%observer's perspective, as follows.

In the following, I give a personalised observers perspective on the 
requirements of a good accretion theory. As highlighted by (King 2012), 
the viscosity in the accretion disk, parametrized in the SS disk, is not understood 
from basic Physics. Till we have a clear understanding of the
viscosity in the accretion disk, accretion theory will not have predictive power like the stellar structure
theory and hence ``we are still a long way
from a theory of accretion 
discs with real predictive 
Power" (King 2012).
 Second, the innermost region of the accretion disk is difficult to
understand and new observations seem to challenge the prevailing views:
%SS disk makes several simplifying assumptions like gravitational energy
%getting thermalised at every radius, hence assuming steady state which may not be %applicable at the
%inner most regions. Further, the inner boundary condition and the transition of the SS %disk into
%other forms of disk (slim disk, advection dominated, etc.) are not completely understood.
% In summery, though most of the observational features can be identified
%with some emission regions, a new observation, sometimes requires some new 
%theoretical perspective. In particular, the exact nature of the innermost region of the
%accretion disk and its structure is still an  unexplained  detail. 
for example the peculiar variability characteristics of GRS~1915+105 is possibly due to the very high accretion rate
and this concept is challenged by the discovery of IGR J17091-3624 showing a behaviour similar to that of GRS~1915+105,
but with a vastly lower luminosity (Altamirano et al. 2011; Rao \& Vadawale 2012; 
Pahari et al. 2012).
The mechanism of superluminal ejection (whether it is a series of events or one large emission)
is  unknown and the exact energy source is still an ongoing debate (King et al. 2013).

\subsection{Mass estimates based on scaling laws}

%  One of the most successful observational To make a real progress in our understanding %of the accretion onto  black holes, first
%we need to establish the paradigms which could have some real predictive powers.
One of the most fundamental parameter of the black hole is its mass
and they are measured with a reasonable accuracy. In Figure 1 I  have plotted the accuracy
in the mass measurements available for 39 AGNs and 17 XRBs.
It can be seen that a typical accuracy 
of 
    $\sim$30\% for AGNs and $\sim$10\% for X-ray binaries are achieved. 
One of the ways to have a clear paradigm of accretion theory is to predict the
mass of the black hole using the observations coming close to the black hole:
this will clearly establish the basic boundary conditions.
%I demonstrate 
%here that the current understanding of accretion theory and the scaling laws do not
%have enough predictive powers.
In Figure 2, left panel shows the estimated  mass for AGNs simply by assuming that the
X-ray luminosity (L$_X$) is proportional to the Eddington luminosity (L$_{Edd}$). The 
bottom-left 
panel uses the fundamental plane of black hole activity tying  L$_X$ and radio luminosity 
to the black hole mass (Merloni et al. 2003).
The measured X-ry spectral index is found to be correlated to L$_X$ and this relation is used
as a proxy to measure mass by Gliozzi et al. (2011) and this result is given in the top right
panel.  
  McHardy et al. (2006) derived a tight correlation between the variability time scale,
bolometric luminosity and the mass of the black hole and this relation is used to predict
the mass of black holes in the bottom right panel of Figure 2 (see  Gonzalez-Martin \& Vaughan 2012).
As can be seen from the figure, these scaling laws
are  unable to predict the mass of the black holes. The best achieved, so far, is the break time scale
relation of McHardy et al. (2006) and that too at 75\% rms accuracy.
Hence, to establish the accretion disk paradigms, it is necessary to evolve
a clear theoretical basis which can clearly provide a definite relation between 
the observables (X-ray luminosity, timing features, spectral features etc.)
and the most fundamental feature of the black hole, its mass. 
%If, for example,
%there is a definite length scale in the accretion disk (as argued in the CENBOL
%model), the observed QPO frequency can be related to this length scale and
%using the simultaneously determined spectral parameters, it should be
%possible to identify the energy scale and hence the mass of the black hole. 
%This type of definitive measurements has the potential of pin-pointing all the
%boundary conditions for the accretion disk formalisms.

\begin{figure}
  \centerline{\includegraphics[angle=0,width=4cm]{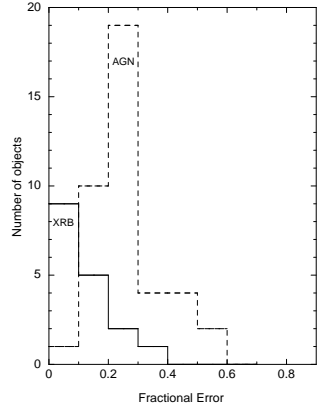}}
  \caption{A histogram of fractional errors in mass for black hole
sources in X-ray binaries and AGNs.}
  \end{figure}

\begin{figure}
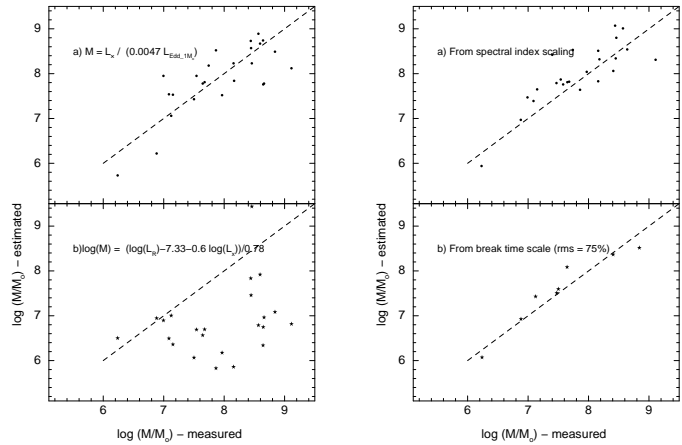

\centerline{\includegraphics[width=4cm]{rad_xray_mass_log.ps} \qquad
            \includegraphics[width=4cm]{Lev_mchardy.ps}}
\caption{The observed and estimated  masses for AGNs (see text).
\label{f:many}}
\end{figure}

\section{Hard X-ray Observations using Astrosat}

Our current understanding of the accretion onto black holes primarily is
the result of the vast amount of data available at the low energies. The 
radiation mechanism, however, could be understood by making a detailed spectral
study in the hard X-rays. This is inherently a difficult field because of the
difficulties of hard X-ray astronomy like lack of source photons,
high background etc. 

For example, measurement of the polarisation in the hard X-rays can be 
used to constrain the emission mechanism. In Cygnus X-1, Laurent et al. (2011)
detected strong polarisation above 250 keV and weak polarisation
below that, thus hinting at different components in the spectrum. This result, however,
does not agree with the overall spectral energy distribution of Cygnus X-1
(Zdziarski et al. 2012), thus highlighting the difficulty of measuring hard X-ray polarisation.
Nustar uses hard X-ray focusing optics and the Nustar data is used to
identify high spin for the supermassive black hole at the centre of NGC~1365
(Risaliti et al., 2013). It probably highlights the difficulties
in the hard X-ray spectral measurements that this result is indeed amenable
to an alternate explanation without requiring high spin for the black hole
(Miller \& Turner 2013). In this context, the forthcoming Astrosat satellite is 
extremely topical for the study of accretion physics.

%\subsection{Astrosat}

Astrosat is an observatory class  satellite dedicated for multi-wavelength observations
(Agrawal 2006). It contains five dedicated instruments and wide band X-ray spectroscopic
observations are done by the co-aligned X-ray instruments: Soft X-ray 
Telescope (SXT), Large Area Xenon-filled Proportional Counter (LAXPC) and
Cadmium Zinc Telluride Imager (CZTI). The Sky Survey Monitor (SSM) provides
a continuous record of the X-ray sky (Seetha et al 2006) and the Ultra-violet Imaging Telescope (UVIT)
makes simultaneous optical and ultra-violet observations (Kumar et al. 2012). 
There are a few very important observational
features which make Astrosat an unique laboratory to understand accretion onto black holes.

LAXPC has the largest ever area above 10 keV and it will extend the RXTE timing capabilities
to higher energies. The SXT can routinely monitor bright sources (Kothare et al. 2009) and hence the crucial
high spectral resolution sensitive measurements at low energies are available for bright
transients. The  LAXPC and CZT-I have individual photon handling capability which is very
essential for the understanding of the systematic in the data (Rao et al. 2010). 
The very low inclination ($\sim$8$^\circ$) orbit ensures very low cosmic ray induced background.
The flexibility to  change/ adjust 
      observation time of  
      SSM pointing is very crucial to track black hole transients.

\begin{figure}
  \centerline{\includegraphics[angle=0,width=4cm]{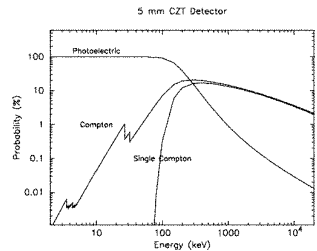}}
  \caption{The absorption probability in the 5 mm CZT detector in the CZT-Imager payload
of Astrosat. The Compton scattering probability is separately shown.}
  \end{figure}

The CZT-Imager has large area (1000 cm$^2$), good energy resolution ($\sim$5\%),
and individual pixel handling and calibration facility. A coded aperture mask is 
used for simultaneous background measurement and identifying multiple sources
in the field of view ($\sim$6$^\circ$). It acts as a pointed collimated detector
upto about 80 -- 100 keV and   
an open all sky monitor above this energy. In conjunction with SSM it will provide 
crucial spectral state monitoring of bright black hole sources. A continuous simultaneous
record in soft and hard X-ray sources for bright ($>$100 mCrab) sources will prove
very useful for a detailed study of state transitions in bright X-ray transients.
The CZT-I also has individual photon counting capability and very good relative time-tagging
capability (20 $\mu$s). The Compton scattering probability is shown in Figure 3. 
It can be seen that there is sufficient probability to detect double events
and these can be identified by the precise time-tagging.
With a pixel size of 2.5 mm, there is sufficient sensitivity to 
measure the azimuthal distribution of the Compton scattered photons.
This results in  polarisation sensitivity above 150 keV
and it can measure Crab polarisation in about a day.
Observations using  Astrosat have the potential to
establish the basic paradigms of accretion onto black holes:\\
%\begin{enumerate}
%\item 
{\bf X-raying the birth of jets:} 
%The exact accretion disk phenomenon responsible for the superluminal jet emission is still unknown. 
In Astrosat, the SSM has flexible observing 
capability which will  provide, to a reasonable accuracy, the possible time of ejection
of a superluminal jet. A coordinated observation, lasting for a few days, with Astrosat 
and other ground based observatories like GMRT, will provide a very detailed and clear 
picture of the accretion disk emission properties responsible for
the jet emission.\\
%\item 
{\bf Disk/ jet symbiosis:} It is still an ongoing debate about the contribution 
of base of the jet for the X-ray spectrum. The CZTI payload can measure polarisation
in bright black hole sources like Cygnus X-1 which will help in a clear segregation
of the disk and jet emission in black hole binaries.\\
%\item 
{\bf An atlas of spectra of black hole states:} 
%Though the RXTE archive has a vast 
%amount of observations, the spectral quality of RXTE is quite poor and hence 
%it is difficult to have a detailed spectral measurements. 
Astrosat
has hard X-ray observations simultaneous   with SXT along with  UV and optical
observations. This will provide a vast amount of good quality spectral and timing
data for at least a dozen black hole sources in different spectral states.\\
%\item 
{\bf Finding new black holes:} In the course of the life time of Astrosat (3 - 5 years), it is expected to discover at least half a dozen new X-ray transients and it will help  in increasing the number of known black hole sources.\\
%\item 
{\bf Finding faint black hole transients:} Pointing Astrosat near the Galactic centre in a 
scan mode will give a sensitive search for faint black hole sources. With the wide filed of
view of CZTI and the superior sensitivity of SXT, the sources discovered by LAXPC can be
quickly identified and faint black hole sources like IGR  J17091-3624 could be
discovered. 
%This will enormously help studying accretion phenomena at vastly
%different dynamic range of X-ray emission.
%\end{enumerate}

In conclusion, black holes in X-ray binaries is  a firmly established paradigm.
The general concepts of accretion are  known reasonably well but 
establishing a firm paradigm requires predictability and better observations,
particularly in the hard X-ray range.

%\section*{Acknowledgements}

%------------------------------------------------------------------------------%
% bibliography: produced from ADS using custom format of                       %
%                                                                              %
%     %z132 \\bibitem[%\2%(y)%\3m]%{R}\n   %\8.1g,%\Y,%\q,%\V,%\ p             %
%------------------------------------------------------------------------------%

\appendix
%------------------------------------------------------------------------------%
% appendices:                                                                  %
%------------------------------------------------------------------------------%

\label{lastpage}
%------------------------------------------------------------------------------%
\end{document}